\documentclass[12pt]{article}
\usepackage{amsmath}
\usepackage{amssymb}
\usepackage{amsfonts}
\usepackage{graphicx}

\newif{\ifcomentarios}
\comentariosfalse

\newtheorem{theorem}{Theorem}

\newtheorem{corollary}[theorem]{Corollary}

\newtheorem{definition}[theorem]{Definition}

\newtheorem{proposition}[theorem]{Proposition}
\newtheorem{remark}[theorem]{Remark}

\begin{document}

\author{Domingos H. U. Marchetti\thanks{%
Present Address: Mathematics Department, \ The University of British
Columbia, Vancouver, \ BC, \ Canada \ V6T 1Z2. Email: \texttt{%
marchett@math.ubc.ca}} \ \ \& \ Walter F. Wreszinski\thanks{%
Email:\texttt{\ wreszins@fma.if.usp.br}} \\
Instituto de F\'{\i}sica\\
Universidade de S\~{a}o Paulo \\
Caixa Postal 66318\\
05314-970 S\~{a}o Paulo, SP, Brasil }
\title{Anderson--like Transition for a Class of Random Sparse Models in $%
d\geq 2$ Dimensions}
\date{}
\maketitle

\begin{abstract}
We show that the Kronecker sum of $d\geq 2$ copies of a random
one--dimensional sparse model displays a spectral transition of the type
predicted by Anderson, from absolutely continuous around the center of the
band to pure point around the boundaries. Possible applications to physics
and open problems are discussed briefly.
\end{abstract}

\section{Introduction and Summary\label{IS}}

\setcounter{equation}{0} \setcounter{theorem}{0}

In this paper we study a class of models whose relationship to the original
Anderson \cite{An} model will now be briefly explained (for further
clarification, see section 3). The Anderson Hamiltonian 
\begin{equation}
H^{\omega }=\Delta +\lambda V^{\omega }\   \label{H}
\end{equation}%
on 
\begin{equation*}
l^{2}(\mathbb{Z}^{d})=\left\{ u=\left( u_{n}\right) _{n\in \mathbb{Z}%
^{d}}:u_{n}\in \mathbb{C},\ \sum_{n\in \mathbb{Z}^{d}}\left\vert
u_{n}\right\vert ^{2}<\infty \right\} ~,\qquad d\geq 1,
\end{equation*}%
is given by the (centered) discrete Laplacian 
\begin{equation}
\left( \Delta u\right) _{n}=\sum_{n^{\prime }:\left\vert n-n^{\prime
}\right\vert =1}u_{n^{\prime }}  \label{laplacian}
\end{equation}%
plus a perturbation by a random potential 
\begin{equation*}
\left( V^{\omega }u\right) _{n}=V_{n}^{\omega }u_{n}
\end{equation*}%
where $\left\{ V_{n}^{\omega }\right\} _{n\in \mathbb{Z}^{d}}$ is a family
of independent, identically distributed random variables (i.i.d.r.v.) on the
probability space $\left( \Omega ,\mathcal{B},\mu \right) $, with a common
distribution $F(x)=\mu \left( \left\{ \omega :V_{n}^{\omega }\leq x\right\}
\right) $; $\lambda >0$ is the disorder parameter also called coupling
constant. The spectrum of $H^{\omega }$ is, by the ergodic theorem, almost
surely a nonrandom set $\sigma (H^{\omega })=[-2d,2d]+\lambda \text{supp}dF$%
. Anderson \cite{An} conjectured that there exists a critical coupling
constant $0<\lambda _{c}<\infty $ such that for $\lambda \geq \lambda _{c}$
the spectral measure of (\ref{H}) is pure point (p.p) for $\mu $--almost
every $\omega $, while, for $\lambda <\lambda _{c}$ the spectral measure of $%
H^{\omega }$ contains two components, separated by so called
\textquotedblleft mobility edge" $E^{\pm }$: if $E\in \lbrack E^{-},E^{+}]$
the spectrum of $H^{\omega }$ is pure absolutely continuous (a.c); in the
complementary set $\sigma (H^{\omega })\backslash \lbrack E^{-},E^{+}]$, $%
H^{\omega }$ has pure point spectra. We refer to \cite{Ji} for a
comprehensive review on the status of the problem and references, and only
wish to remark that for $d=1$ the spectrum is p.p. for all $\lambda $ for
almost every $\omega $ (\cite{GMP,KS}), while, for $d\geq 2$ the existence
of a.c. spectrum is open, except for the version of (\ref{H}) on the Bethe
lattice, where it was first proved by A. Klein in a seminal paper \cite{Kl}
(see also \cite{Ji}, Section 2.31).

Given the above mentioned difficulties, one might be led to study the limit $%
\lambda \rightarrow 0$ of (\ref{H}), for which the spectrum is pure a.c.. We
shall instead follow a different approach to the Anderson conjecture
suggested by Molchanov: the limit of zero concentration, i.e., taking $%
V^{\omega }$ in (\ref{H}) such that%
\begin{equation}
V_{n}^{\omega }=\sum_{i}\varphi _{i}^{\omega }(n-a_{i})~,  \label{V}
\end{equation}%
with elementary potential (\textquotedblleft bump") $\varphi ^{\omega }:%
\mathbb{Z}^{d}\longrightarrow \mathbb{R}$ satisfying a uniform integrability
condition%
\begin{equation}
\left\vert \varphi ^{\omega }(z)\right\vert \leq \frac{C_{0}}{1+\left\vert
z\right\vert ^{d+\varepsilon }}  \label{phi}
\end{equation}%
for some $\varepsilon >0$ and $0<C_{0}<\infty $ and%
\begin{equation}
\lim_{R\rightarrow \infty }\frac{\#\left\{ i:\left\vert a_{i}\right\vert
\leq R\right\} }{R^{d}}=0\ .  \label{a}
\end{equation}%
Due to condition (\ref{a}) of zero concentration, potentials such as (\ref{V}%
) are called sparse and have been intensively studied in recent years since
the seminal work by Pearson in dimension $d=1$ \cite{Pe}, notably by
Kiselev, Last and Simon \cite{KLS} for $d=1$ and by Molchanov in the
multidimensional case \cite{Mo1} (see also \cite{MoV, Mo2} for complete
proofs and additional results). As a consequence of (\ref{phi}), for $d\geq 2
$ the interaction between bumps is weak \cite{Mo1} while for $d=1$ the phase
of the wave after propagation between distant bumps become \textquotedblleft
stochastic" \cite{Pe}. This is the right moment to introduce our
one--dimensional model.

Instead of (\ref{H}) we shall adopt an off--diagonal Hamiltonian which
contains the Laplacian (\ref{laplacian}): 
\begin{equation}
J^{\omega }\equiv J_{P^{\omega }}=\left( 
\begin{array}{ccccc}
0 & p_{0} & 0 & 0 & \cdots \\ 
p_{0} & 0 & p_{1} & 0 & \cdots \\ 
0 & p_{1} & 0 & p_{2} & \cdots \\ 
0 & 0 & p_{2} & 0 & \cdots \\ 
\vdots & \vdots & \vdots & \vdots & \ddots%
\end{array}%
\right) \;,  \label{J}
\end{equation}%
\noindent for each sequence $P^{\omega }=(p_{n}^{\omega })_{n\geq 0}$ of the
form%
\begin{equation}
p_{n}^{\omega }=\left\{ 
\begin{array}{lll}
p & \mathrm{if} & n=a_{j}^{\omega }~\text{for some }j \\ 
1 & \mathrm{if} & \text{otherwise}\,,%
\end{array}%
\right.  \label{p}
\end{equation}%
\noindent for $p\in (0,1)$. Above, $\{a_{j}^{\omega }\}_{j\geq 1}$ is a
random set of natural numbers 
\begin{equation*}
a_{j}^{\omega }=a_{j}+\omega _{j}
\end{equation*}%
with $a_{j}$ satisfying the \textquotedblleft sparseness" condition 
\begin{equation}
a_{j}-a_{j-1}=\beta ^{j}\;,\qquad \qquad j=2,3,\ldots  \label{sparse}
\end{equation}%
\noindent with $a_{1}+1=\beta \geq 2$ where $\beta $ is an integer and $%
\omega _{j}$, $j\geq 1$, are independent random variables defined on a
probability space $(\Omega ,\mathcal{B},\nu )$, uniformly distributed on the
set $\Lambda _{j}=\{-j,\ldots ,j\}$. We denote by $J_{\phi }^{\omega }$ an
operator related to the Jacobi matrix $J^{\omega }$ acting on the Hilbert
space $\mathcal{H}$ of square summable complex valued sequences $u=\left(
u_{n}\right) _{n\geq -1}$ satisfying a $\phi $-boundary condition at $-1$:%
\begin{equation}
\left( J_{\phi }^{\omega }u\right) _{n}=p_{n-1}^{\omega }u_{n-1}+p_{n}u_{n+1}
\label{Ju}
\end{equation}%
for $n\geq 0$, with $p_{-1}^{\omega }=1$ and 
\begin{equation}
u_{-1}\cos \phi -u_{0}\sin \phi =0  \label{phicond}
\end{equation}%
(i.e., $\left( J_{\phi }^{\omega }u\right) _{n}=\left( J_{0}^{\omega
}u\right) _{n}+\delta _{0,n}\tan \phi ~u_{0}$). The variables $\left\{
\omega _{j}\right\} _{j\geq 1}$ introduce uncertainty in the positions $%
\{a_{j}\}_{j\geq 1}$ where the \textquotedblleft bumps" are located. The
corresponding diagonal version satisfies trivially (\ref{phi}), since $%
\varphi _{i}^{\omega }(n)=\delta _{\omega _{i},n}$ is just a Kronecker delta
at $\omega _{i}$; such models are nowadays called Poisson models (see pg.
624 of \cite{Ji} and references therein). A disordered diagonal model of the
above type- to which our results are also applicable- was introduced by Zlato%
\v{s} \cite{Zl}. The present non--diagonal version has some advantages in
addition to the initial motivation coming from \cite{H1}: that the spectrum $%
\sigma (J^{\omega })$ of $J^{\omega }$ interpolates between purely
absolutely continuous for $p=1$ and dense pure point for $p=0$ (in the
latter case, $J^{\omega }$ is a direct sum of finite matrices; the dense
character is due to (\ref{sparse})). It is easily proved that the essential
spectrum of $J^{\omega }$ is $\sigma _{\mathrm{ess}}(J^{\omega })=[-2,2]$
(see \cite{CMW1}).

We may ask whether the p.p. part of $\sigma (J^{\omega })$ for $p=0$ above
persists in some nonempty interval. Let%
\begin{equation}
I=\left\{ \lambda \in \lbrack -2,2]:v^{-2}(\beta -1)(4-\lambda ^{2})\geq
1\right\}  \label{I}
\end{equation}%
where $v=v(p)=(1-p^{2})/p$ \noindent and set%
\begin{equation}
I^{c}=[-2,2]\backslash I~.  \label{Ic}
\end{equation}%
Note that $I=\emptyset $ (consequently, $I^{c}=\left[ -2,2\right] $) if $%
p<p_{c}$, where $p_{c}$ is defined by%
\begin{equation*}
v^{2}(p_{c})=\left( \frac{1-p_{c}^{2}}{p_{c}}\right) ^{2}=4(\beta -1)~.
\end{equation*}%
Such a equation has always a solution $p_{c}=\sqrt{2\beta -1-2\sqrt{\beta
^{2}-\beta }}$ in $\left( 0,1\right) $ for $\beta \geq 2$ and $%
v_{c}=v(p_{c})=2\sqrt{\beta -1}$ will play a role similar to the critical
coupling $\lambda _{c}$ of the Anderson model. We have (see Theorem 2.4 of 
\cite{CMW1})

\begin{theorem}
\label{sharp}Let $J_{\phi }^{\omega }$ be defined by (\ref{J})--(\ref%
{phicond}), and set%
\begin{eqnarray}
A_{\mathrm{sc}} &=&2\cos \pi \mathbb{Q}\cap I  \notag \\
A_{\mathrm{pp}} &=&2\cos \pi \mathbb{Q}\cap I^{c}~.  \label{AA}
\end{eqnarray}%
Then, for $\nu $--almost every $\omega $,

\begin{enumerate}
\item[a.] the spectrum of $J_{\phi }^{\omega }$ restricted to the set $%
I\backslash A_{\mathrm{sc}}^{\prime }$ with $A_{\mathrm{sc}}^{\prime }=A_{%
\mathrm{sc}}\cup A^{\prime }$ and $A^{\prime }$ a set of Lebesgue measure
zero, is purely singular continuous;

\item[b.] the spectrum of $J_{\phi }^{\omega }$ is dense pure point when
restricted to $I^{c}\backslash A_{\mathrm{pp}}$ for almost every $\phi \in
\lbrack 0,\pi )$.
\end{enumerate}
\end{theorem}

\begin{remark}
\label{refine}

\begin{enumerate}
\item The occurrence of the set $A^{\prime }$ of Lebesgue measure zero is
related to the definition of essential (or minimal) support of the spectral
measure $\mu $ (see Definition 1 of \cite{GP})).

\item As we have excluded a countable set $A_{pp}$, the spectrum is purely
p.p. in $I^{c}$.
\end{enumerate}
\end{remark}

\begin{figure}[ht!]
\label{mobility} \centering
\includegraphics[scale=1]{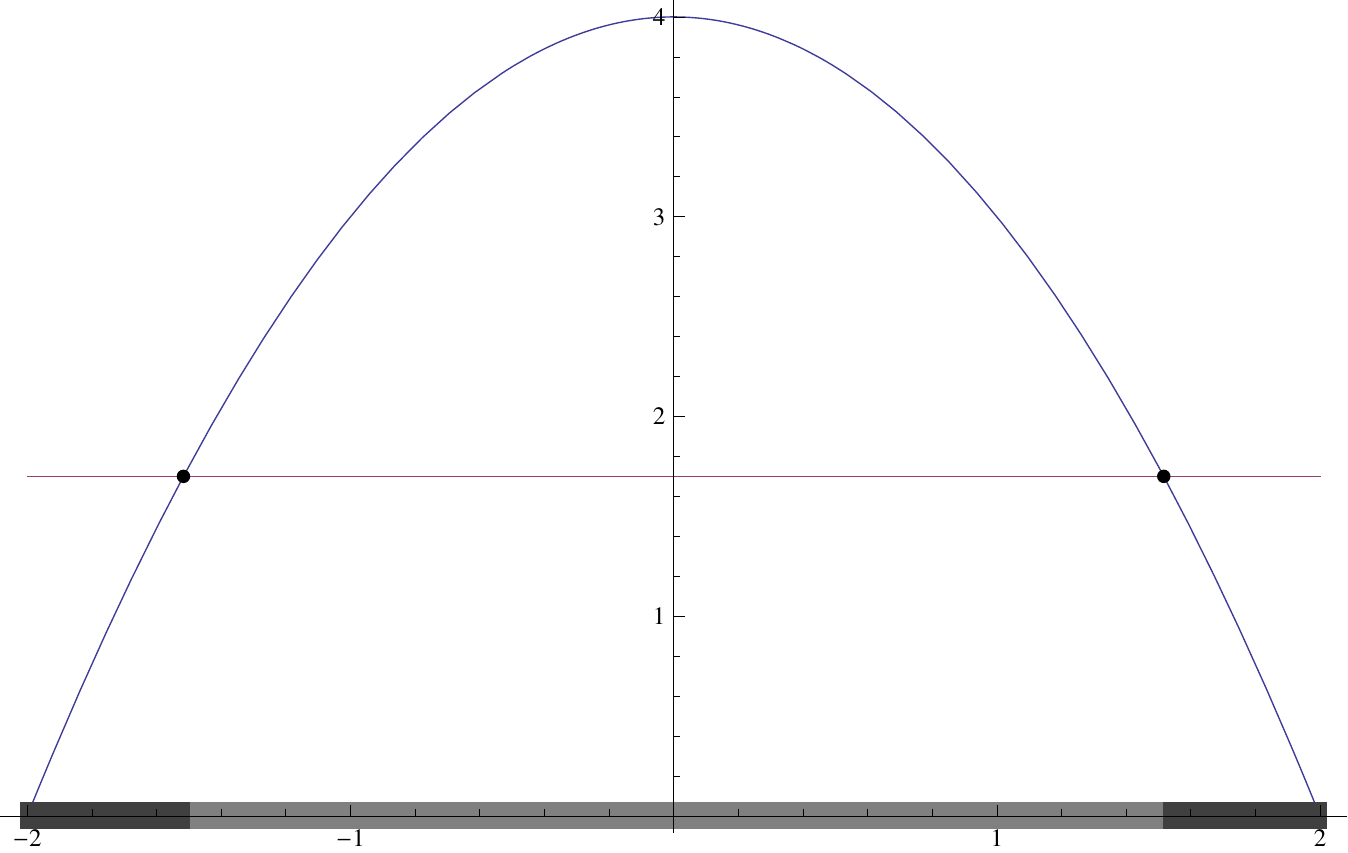}
\caption{Singular continuous (light gray) and pure point (dark gray) spectra
separated by the ``mobility edges'' $\protect\lambda ^{\pm}= \pm 2 \protect%
\sqrt{1 - v^{2}/v^{2}_c}$; $v / v_{c} = 1.3038...$}
\end{figure}

Theorem \ref{sharp} for the corresponding diagonal model was proved in \cite%
{Zl}, except for the specification of the set $A_{pp}$, which leads to the
refinement of Remark \ref{refine}.2. The latter depended on the details of
the method in \cite{CMW1}, whose crucial step was a proof that the sequence
of Pr\"{u}fer angles $\left( \theta _{j}^{\omega }\right) _{j\geq 0}$ (see 
\cite{KLS,Zl,MWGA} for definitions) is uniformly distributed mod $\pi $
(u.d. mod $\pi $) for $\nu $--almost every $\omega $ and for all $\lambda
=2\cos \varphi $ with $\varphi \in \lbrack 0,\pi ]$ such that $\varphi /\pi $
is an irrational number. As remarked by Remling \cite{Re} in his review of \cite{MWGA}%
, which introduced our method, the new idea was to fix the energy $\lambda $
and assume (or prove, when one is able to) that the Pr\"{u}fer angles $%
\left( \theta _{j}^{\omega }\right) $ at $a_{j}$ are uniformly distributed
(u.d.) as a function of $j$, instead of the traditional approach which
exploits the u.d. of the Pr\"{u}fer angles in the energy variable at fixed $%
a_{j}$. We shall see that this refinement, perhaps of apparently minor
importance, will play an important role in our approach (see Remark 2.9).
Figure \ref{mobility} depicts the one--dimensional spectral transition,
where the \textquotedblleft mobility edges" $\lambda ^{\pm }=2\cos \varphi
^{\pm }$ are implicitly given by the equation 
\begin{equation*}
1-\frac{\lambda ^{2}}{4}=\sin ^{2}\varphi =\frac{v^{2}}{v_{c}^{2}}
\end{equation*}%
provided $v<v_{c}=2\sqrt{\beta -1}$.

For superexponential sparseness, i.e., $a_{j}-a_{j-1}=\left[ e^{cn^{\gamma }}%
\right] $ ($\left[ z\right] $ the integer part of $z$), with $c>0$, $\gamma
>1$ and $\left\{ \omega _{j}\right\} _{j\geq 1}$ independent random
variable, uniform in $\Lambda _{j}$, it may be proved that $\sigma (J_{\phi
}^{\omega })$ is purely singular continuous (s.c.) for almost every $\omega
\in \times _{j=1}^{\infty }$ $\{-j,\ldots ,j\}$ (\cite{CMW1}, Theorem 5.2).
This has a simple physical interpretation already pointed out by Pearson 
\cite{Pe}: the enormous separation between the $a_{j}$ causes the
aforementioned \textquotedblleft stochasticity" of the phase of the Bloch
wave of difference Laplacian , with the particle behaving as if successively
undergoing reflections (and transmissions) through the bumps. The reflection
from the latter is $O\left( v^{2}\right) $ by the Born approximation, and,
since $\displaystyle\sum_{n\geq 0}(1-p_{n}^{2})^{2}/p_{n}^{2}=\infty $, no
particles arrive at infinity (for $\displaystyle\sum_{n\geq
0}(1-p_{n}^{2})^{2}/p_{n}^{2}<\infty $, the spectrum is purely a.c. as may
be proved by methods of \cite{KLS}). This conclusion is rigorously confirmed
by the dynamics: the average time spent by the particle, in any bounded
region, is zero for states both in the a.c. and s.c. subspaces, by the RAGE
theorem \cite{RS}, but the \textquotedblleft sojourn time" (properly
defined, see \cite{Si}) for a particle in the s.c. subspace has, in contrast
to the a.c. case, to be infinite for some finite region of space as a
consequence of Theorem 1 of \cite{Si}.

On the other hand, for subexponential sparseness, with $a_{j}-a_{j-1}=\left[
e^{cn^{\gamma }}\right] $ with $\gamma <1$ and everything else as before, $%
\sigma _{\mathrm{ess}}(J_{\phi }^{\omega })=\sigma _{\mathrm{pp}}(J_{\phi
}^{\omega })=\left[ -2,2\right] $ for a.e. boundary phase $\phi \in \left[
0,\pi \right] $ and for a.e. $\omega \in \times _{j=1}^{\infty }$ $%
\{-j,\ldots ,j\}$ (\cite{CMW1}, Theorem 5.1).

These results joins smoothly to the one (corresponding to $\gamma =0$) for
the standard Anderson model in $d=1$, according to which all states are
localized \cite{GMP,KS}. The latter is believed to be physically related to
the subtle instability of tunneling \cite{JMS,S1} which is strongest in $d=1$%
.

What is really surprising in Theorem \ref{sharp} is, of course, not the
existence of s.c. spectrum, but that of p.p. spectrum in a regime of high
(exponential) sparsity (\ref{sparse}). That is the more so because the
well--known instability of Anderson localization under rank one
perturbations \cite{dR} implies that the spectral measure associated to s.c.
spectrum which is obtained in the Anderson model by changing the value of
the potential at a point is supported on a set of zero Hausdorff dimension,
which is not the case for $J_{\phi }^{\omega }$ (see \cite{Zl,CMW2}). Thus
the spectral transition depicted in the latter is of the robust type. For
further general references on random systems, see \cite{CL}, \cite{PF}, \cite%
{Sto}.

We now summarize the contents of the paper. In Section \ref{MR} we prove our
main result (Theorem \ref{main}), which states that the Kronecker sum of $%
d\geq 2$ copies of $J_{\phi }^{\omega }$ exhibits a Anderson transition (see
also Section \ref{COP} for this designation and a discussion of possible
application to the Anderson transition in lightly-doped semiconductors) from
a.c. spectrum for small energy (i.e., in the region situated around the
center of the band) to dense p.p. for large energy (i.e., in the union of
the two regions around the extreme points): this is true for suitable values
of parameters, and exclusion of resonances.

The proof of our main result (Theorem \ref{main}) shows that ideas of Kahane
and Salem \cite{KS1,KS2} combine with the Strichartz-Last theorem \cite%
{Str,L1} in a neat way, yielding a result of quite general nature, i.e.,
showing the existence of a.c. spectrum for any Kronecker sum of operators $%
A\otimes I+\theta I\otimes A$ for a.e. $\theta \in \lbrack 0,1]$ whenever $A$
has s.c. spectrum in some nonempty interval with local Hausdorff dimension
greater than $1/2$. For this reason, we believe that the idea might have
further potential applications, e.g., to the intermediate region, see the
discussion in Section \ref{COP}. For a physically related model - the
Anderson random potential on tree graphs (i.e. Bethe lattice) at weak
disorder, absence of mobility edge has been shown recently \cite{AW1}. We
also refer to \cite{AW2} for the important proof of existence of a.c.
spectra in quantum tree graphs with weak disorder, as well as \cite{AW1} for
further literature on quantum tree graphs.

\section{Main Result\label{MR}}

\setcounter{equation}{0} \setcounter{theorem}{0}

In order to formulate and prove our main result (Theorem \ref{main}) we need
the following \cite{Zl}:

\begin{definition}
\label{exact}A finite Borel measure $\mu $ has exact local Hausdorff
dimension $\alpha (\cdot )$ in an interval $I$ if for any $\lambda \in I$
there exists an $\alpha (\lambda )$ such that for any $\varepsilon >0$ there
is a $\delta >0$ with $\mu \left( (\lambda -\delta ,\lambda +\delta )\cap
\cdot \right) $ is both $(\alpha (\lambda )-\varepsilon )$--continuous and $%
(\alpha (\lambda )+\varepsilon )$--singular.
\end{definition}

The above notion of continuous and singular refer to the Hausdorff measure $%
h^{\alpha }$ (see e.g. Section 4 of \cite{L1} for a convenient summary of
all relevant concepts and references).

\begin{definition}[Definition 2.1 of \protect\cite{L1}]
\label{holder}We say that $\mu $ is uniformly $\alpha $--H\"{o}lder
continuous (U$\alpha $H) iff there exists a constant $C$ such that, for
every interval $I$ with $\left\vert I\right\vert <1$, 
\begin{equation*}
\mu \left( I\right) <C\left\vert I\right\vert ^{\alpha }
\end{equation*}
\end{definition}

Above, $\left\vert S\right\vert $ denotes Lebesgue measure of $S$. Let $%
\left\{ E(\lambda )\right\} $ denote the spectral family associated to $%
J_{\phi }^{\omega }$ (we omit the indices for simplicity) and $\left\{ E_{%
\mathrm{sc}}(\lambda )\right\} $, $\left\{ E_{\mathrm{pp}}(\lambda )\right\} 
$ its singular continuous and pure point parts. As usual (see e.g. \cite{KS}%
), we define $\mathcal{H}_{\mathrm{sc}}$ and $\mathcal{H}_{\mathrm{pp}}$ so
that, if $\psi \in \mathcal{H}_{sc}$ the spectral measure, 
\begin{subequations}
\begin{equation}
\mu _{\psi }^{\mathrm{sc}}(\lambda )\equiv \left( \psi ,E(\lambda )\psi
\right) ~,  \label{mu1}
\end{equation}%
is purely singular continuous and, if $\psi \in \mathcal{H}_{\mathrm{pp}}$, 
\begin{equation}
\ \mu _{\psi }^{\mathrm{pp}}(\lambda )\equiv \left( \psi ,E(\lambda )\psi
\right) ~,  \label{mu2}
\end{equation}%
is purely pure point. $\mathcal{H}_{\mathrm{sc}}$ and $\mathcal{H}_{\mathrm{%
pp}}$ are closed (in norm), mutually orthogonal subspaces: $\mathcal{H=H}_{%
\mathrm{sc}}\oplus \mathcal{H}_{\mathrm{pp}}$, and invariant under $J_{\phi
}^{\omega }$.

By \cite{Zl,CMW2} the local Hausdorff dimension (Definition \ref{exact})
associated to $J_{\phi }^{\omega }\upharpoonright I$, with $I$ given by (\ref%
{I}), is 
\end{subequations}
\begin{equation}
\alpha (\lambda )=1-\frac{\log r(\lambda )}{\log \beta }  \label{alpha}
\end{equation}%
where%
\begin{equation}
r(\lambda )=1+\frac{v^{2}}{4-\lambda ^{2}}~.  \label{r}
\end{equation}%
We now choose an arbitrary $\varepsilon >0$ and pick $\left( \lambda
_{i}\right) _{i=1}^{N_{\varepsilon }}$ with $\lambda _{i}\in I$ and $\left(
\delta _{\varepsilon }^{i}\right) _{i=1}^{N_{\varepsilon }}$, with 
\begin{subequations}
\label{d}
\begin{equation}
0<\delta _{\varepsilon }^{i}<1~,  \label{delta}
\end{equation}%
for some $N_{\varepsilon }<\infty $, in such way that 
\begin{eqnarray}
\lambda _{1}-\delta _{\varepsilon }^{1} &=&-\sqrt{4-v^{2}/(\beta -1)}~,
\label{delta-a} \\
\lambda _{i}+\delta _{\varepsilon }^{i} &=&\lambda _{i+1}-\delta
_{\varepsilon }^{i+1}\ ,\quad i=1,\ldots ,N_{\varepsilon }-1~,
\label{delta-b} \\
\lambda _{N_{\varepsilon }}+\delta _{\varepsilon }^{N_{\varepsilon }} &=&%
\sqrt{4-v^{2}/(\beta -1)}~.  \label{delta-c}
\end{eqnarray}%
We set%
\begin{equation}
A_{\varepsilon }^{i}=[\lambda _{i}-\delta _{\varepsilon }^{i},\lambda
_{i}+\delta _{\varepsilon }^{i})~,  \label{delta-d}
\end{equation}%
for $1\leq i<N_{\varepsilon }$, with $A_{\varepsilon }^{N_{\varepsilon
}}=[\lambda _{N_{\varepsilon }}-\delta _{\varepsilon }^{N_{\varepsilon
}},\lambda _{N_{\varepsilon }}+\delta _{\varepsilon }^{N_{\varepsilon }}]$,
and 
\begin{equation}
~\tilde{A}_{\varepsilon }^{i}=(\lambda _{i}-\delta _{\varepsilon
}^{i},\lambda _{i}+\delta _{\varepsilon }^{i})~,  \label{delta-e}
\end{equation}%
for $1\leq i\leq N_{\varepsilon }$, and write $I$ as a mutually disjoint
union: 
\end{subequations}
\begin{equation}
I=\bigcup_{i=1}^{N_{\varepsilon }}A_{\varepsilon }^{i}~.  \label{delta-f}
\end{equation}%
Observe that (\ref{delta-a}) and (\ref{delta-c}) represent the boundary
points $\lambda _{\pm }$ of $I$, given by (\ref{I}). The choice of $\left(
\lambda _{i}\right) _{i=1}^{N_{\varepsilon }}$ is arbitrary but the
quantities $\delta _{\varepsilon }^{i}$, $i=1,\ldots ,N_{\varepsilon }$, are
chosen in correspondence to $\varepsilon $ according to Definition \ref%
{exact}, with $\alpha (\cdot )$ given by (\ref{alpha}), and satisfy 
\begin{equation}
\bar{\delta}_{\varepsilon }\equiv \max_{i}\delta _{\varepsilon
}^{i}\rightarrow 0~,  \label{delta-bar}
\end{equation}%
by continuity, as $\varepsilon $ tends to $0$. As a consequence, the
spectral measure of $J_{\phi }^{\omega }$ restricted to $\tilde{A}%
_{\varepsilon }^{i}$ 
\begin{equation}
\mu _{\psi }^{\mathrm{sc}}\upharpoonright \tilde{A}_{\varepsilon }^{i}
\label{restrict}
\end{equation}
is $\left( \alpha(\lambda _{i})-\varepsilon \right) $--continuous and $%
\left( \alpha (\lambda_{i})+\varepsilon \right) $--singular, for $i=1,\ldots
,N_{\varepsilon }$.

\begin{proposition}
\label{dense}Under the hypotheses of Theorem \ref{sharp} and (\ref{alpha})--(%
\ref{restrict}), there exists a dense set $D$ in $\mathcal{H}_{\mathrm{sc}}$
such that, $\forall \psi \in D$, $\mu _{\psi }^{\mathrm{sc}}\upharpoonright 
\tilde{A}_{\varepsilon }^{i}$ is, for each $i\in \left\{ 1,,\ldots
,N_{\varepsilon }\right\} $, uniformly $\left( \alpha (\lambda
_{i})-\varepsilon \right) $--H\"{o}lder continuous.
\end{proposition}

\noindent \textit{Proof.} We write%
\begin{equation*}
\mathcal{H}=\displaystyle\bigoplus_{i=1}^{N_{\varepsilon }}\mathcal{H}_{i}
\end{equation*}%
where $\mathcal{H}_{i}$ is the subspace of $\mathcal{H}_{\mathrm{sc}}$
generated by 
\begin{equation*}
\left\{ E_{I}\psi :\psi \in \mathcal{H}_{\mathrm{sc}}~,\ \text{for every }%
I=(\lambda ,\lambda ^{\prime }]\subset \tilde{A}_{\varepsilon }^{i}\right\}
\end{equation*}%
where $E_{I}=\displaystyle\int_{I}dE(\lambda )$ is the spectral projection
on $I$. By (\ref{alpha})--(\ref{restrict}) and Theorem 5.2 of \cite{L1}, for
each $\mathcal{H}_{i}$ we may choose $D_{i}$ dense in $\mathcal{H}_{i}$ such
that, $\forall \psi \in D_{i}$, $\mu _{\psi }$ is uniformly $\left( \alpha
(\lambda _{i})-\varepsilon \right) $--H\"{o}lder continuous. Since the
subspace $\mathcal{M}$ generated by $\left\{ E(\lambda _{i}+\delta
_{\varepsilon }^{i})\psi :\psi \in \mathcal{H}\right\} $ for $i=1,\ldots
,N_{\varepsilon }-1$ is such that $\mathcal{M}\subset \mathcal{H}_{\mathrm{sc%
}}^{\bot }$, we have by (2.4c), (2.4e) and (2.5) that $\bigoplus_{i=1}^{N_{%
\varepsilon }}D_{i}$ is dense in $\mathcal{H}_{\mathrm{sc}}$ and satisfies
the assertion by (\ref{restrict}).

\hfill $\Box $

\begin{corollary}
\label{corl}Let $I_{0}\subseteq I$ and $\psi \in D$. Then $\mu _{\psi }^{%
\mathrm{sc}}\upharpoonright I_{0}$ is U$\alpha $H, where%
\begin{equation}
\alpha =\min_{i:\tilde{A}_{\varepsilon }^{i}\cap I_{0}\neq \emptyset }\alpha
(\lambda _{i})-\varepsilon ~.  \label{alpha-def}
\end{equation}
\end{corollary}

\noindent \textit{Proof.} This follows immediately from Proposition \ref%
{dense}, Definition \ref{holder} and additivity of $\mu _{\psi }^{\mathrm{sc}%
}$.

\hfill $\Box $

In the rest of the paper we assume that $\varepsilon $ and $\left( \delta
_{\varepsilon }^{i}\right) _{i=1}^{N_{\varepsilon }}$ is a given fixed set
of numbers, with $\varepsilon >0$ arbitrarily small (but with $%
N_{\varepsilon }<\infty $). Consider the Kronecker sum of two copies of $%
J_{\phi }^{\omega }$ as an operator on $\mathcal{H} \otimes \mathcal{H}$: 
\begin{equation}
J_{\theta }^{(2)}:=J_{\phi }^{\omega ^{1}}\otimes I+\theta I\otimes J_{\phi
}^{\omega ^{2}}  \label{J2}
\end{equation}%
where $\omega ^{1}=\left( \omega _{j}^{1}\right) _{j\geq 1}$ and $\omega
^{2}=\left( \omega _{j}^{2}\right) _{j\geq 1}$ are two independent sequences
of independent random variables defined in $\left( \Omega ,\mathcal{B},\nu
\right) $, as before (we omit $\omega ^{1}$ and $\omega ^{2}$ in the l.h.s.
of (\ref{J2}) for brevity). Above, the parameter $\theta \in \lbrack 0,1]$
is included to avoid resonances (see Remark \ref{resonance}). We ask for
properties of $J_{\theta }^{(2)}$ (e.g. the spectral type) which hold for 
\textbf{typical} configurations, i.e., a.e. $\left( \omega ^{1},\omega
^{2},\theta \right) $ with respect to $\nu \times \nu \times l$ where $l$ is
the Lebesgue measure in $[0,1]$. $J_{\theta }^{(2)}$ is a special
two--dimensional analog of $J_{\phi }^{\omega }$; if the latter was replaced
by $-\Delta +V$ on $L^{2}(\mathbb{R},dx)$ where $\Delta =d^{2}/dx^{2}$ is
the second derivative operator, and $V$ a multiplicative operator $V\psi
(x)=V(x)\psi (x)$ (potential), the sum (\ref{J2}) would correspond to $%
\left( -d^{2}/dx_{1}^{2}+V_{1}\right) +\left( -d^{2}/dx_{2}^{2}+V_{2}\right) 
$ on $L^{2}(\mathbb{R}^{2},dx_{1}dx_{2})$, i.e., the \textquotedblleft
separable case" in two dimensions. Accordingly, we shall also refer to $%
J_{\theta }^{(n)}$, $n=2,3,\ldots $, as the separable case in $n$ dimensions.

Our approach is to look at the quantity 
\begin{subequations}
\begin{equation}
\left( \Phi ,e^{-itJ_{\theta }^{(2)}}\Psi \right) =f^{1}(t)f^{2}(\theta t)
\label{ff}
\end{equation}%
by (\ref{J2}), where 
\begin{equation}
f^{i}(s)=f_{\mathrm{sc}}^{i}(s)+f_{\mathrm{pp}}^{i}(s)~,\quad i=1,2
\label{f+f}
\end{equation}%
with 
\begin{eqnarray}
f_{\mathrm{sc}}^{i}(s) &=&\int e^{-is\lambda }d\mu _{\varphi _{i},\psi
_{i}}^{\mathrm{sc}}(\lambda )  \label{fsc} \\
f_{\mathrm{pp}}^{i}(s) &=&\int e^{-is\lambda }d\mu _{\rho _{i},\chi _{i}}^{%
\mathrm{pp}}(\lambda )  \label{fpp}
\end{eqnarray}%
Above $\Phi ,\Psi \in \mathcal{H}\otimes \mathcal{H}$, 
\end{subequations}
\begin{subequations}
\begin{eqnarray}
\Phi &=&\left( \varphi _{1}\dot{+}\rho _{1}\right) \otimes \left( \varphi
_{2}\dot{+}\rho _{2}\right) ~,  \label{Phi} \\
\Psi &=&\left( \psi _{1}\dot{+}\chi _{1}\right) \otimes \left( \psi _{2}\dot{%
+}\chi _{2}\right) ~,  \label{Psi}
\end{eqnarray}%
with $\varphi _{i},\psi _{i}\in \mathcal{H}_{\mathrm{sc}}$, $\rho _{i},\chi
_{i}\in \mathcal{H}_{\mathrm{pp}}$ and $\varphi \dot{+}\rho $ denotes the
direct sum of two vectors $\varphi ,\rho \in \mathcal{H}$. The vectors 
\begin{equation}
\varphi _{1},\psi _{1}\in D_{1}\ ,\quad \varphi _{2},\psi _{2}\in D_{2}
\label{phipsi}
\end{equation}%
where $D_{1}$ and $D_{2}$ are copies of the set $D$ occurring in Proposition %
\ref{dense}; by (\ref{Phi}), (\ref{Psi}) and (\ref{phipsi}), $\varphi _{i}%
\dot{+}\rho _{i},~\psi _{i}\dot{+}\chi _{i}$ run through a dense set in $%
\mathcal{H}=$ $\mathcal{H}_{\mathrm{sc}}\oplus \mathcal{H}_{\mathrm{pp}}$.
In (\ref{fsc}) and (\ref{fpp}), $\mu _{\varphi ,\psi }^{\mathrm{sc}}(\lambda
)=\left( \varphi ,E(\lambda )\psi \right) $, $\mu _{\rho ,\chi }^{\mathrm{pp}%
}(\lambda )=\left( \rho ,E(\lambda )\chi \right) $ as in (\ref{mu1}) and (\ref{mu2}), the 
$f$'s being the corresponding Fourier--Stieltjes (F.S.) transforms. By (\ref{ff}) and 
(\ref{f+f}) 
\end{subequations}
\begin{subequations}
\begin{equation}
\left( \Phi ,e^{-itJ_{\theta }^{(2)}}\Psi \right) =g(t,\theta )+h(t,\theta
)+k(t,\theta )  \label{three}
\end{equation}%
where 
\begin{eqnarray}
g(t,\theta ) &=&f_{\mathrm{sc}}^{1}(t)f_{\mathrm{sc}}^{2}(\theta t)
\label{three1} \\
h(t,\theta ) &=&f_{\mathrm{sc}}^{1}(t)f_{\mathrm{pp}}^{2}(\theta t)+f_{%
\mathrm{pp}}^{1}(t)f_{\mathrm{sc}}^{2}(\theta t)  \label{three2} \\
k(t,\theta ) &=&f_{\mathrm{pp}}^{1}(t)f_{\mathrm{pp}}^{2}(\theta t)
\label{three3}
\end{eqnarray}%
are the F.S. transforms of the complex valued spectral measures of $%
J_{\theta }^{(2)}$ associated with $\mathcal{H}_{\mathrm{sc}}\oplus \mathcal{%
H}_{\mathrm{sc}}$, $\mathcal{H}_{\mathrm{sc}}\oplus \mathcal{H}_{\mathrm{pp}%
}\cup \mathcal{H}_{\mathrm{pp}}\oplus \mathcal{H}_{\mathrm{sc}}$ and $%
\mathcal{H}_{\mathrm{pp}}\oplus \mathcal{H}_{\mathrm{pp}}$, respectively. It
follows from (\ref{three1}) that $g$ is F.S. transform of the convolution of
the measures $\mu _{\varphi _{1}}^{\mathrm{sc}}$ and $\tilde{\mu}_{\varphi
_{2}}^{\mathrm{sc}}$ with 
\end{subequations}
\begin{equation}
\tilde{\mu}_{\varphi _{2}}^{\mathrm{sc}}(\lambda )\equiv \mu _{\varphi
_{2}}^{\mathrm{sc}}(\lambda /\theta )~,\qquad \theta \neq 0  \label{mu/theta}
\end{equation}%
defined by (see \cite{Kat}, pg. 41):%
\begin{equation}
\mu _{\varphi _{1}}^{\mathrm{sc}}\ast \tilde{\mu}_{\varphi _{2}}^{\mathrm{sc}%
}(B)=\int \mu _{\varphi _{1}}^{\mathrm{sc}}(B-\lambda )d\tilde{\mu}_{\varphi
_{2}}^{\mathrm{sc}}(\lambda )  \label{conv}
\end{equation}%
for any Borel set $B$ of $\mathbb{R}$, where $B-\lambda \equiv B-\left\{
\lambda \right\} =\left\{ x-\lambda :x\in B\right\} $, and analogously for $%
h $ and $k$.

At least since the paper of Kahane and Salem \cite{KS1} of 1958, it is well
known that the convolution of two s.c. measures may be absolutely continuous
(this possibility was revived for models in mathematical physics by \cite{MM}%
). Their proof, as well as our proof of the corresponding assertion in the
forthcoming Theorem \ref{main}, was based on the following folklore
proposition:

\begin{proposition}
\label{ac}Let $\mu $ be a measure on the space $M(\mathbb{R})$ of all finite
regular Borel measures on $\mathbb{R}$. If the Fourier--Stieltjes transform
of $\mu $%
\begin{equation}
\mathbb{R}\ni t\longmapsto \hat{\mu}(t)=\int e^{-it\lambda }d\mu (\lambda )
\label{fs}
\end{equation}%
belongs to $L^{2}(\mathbb{R},dt)$, then $\mu $ is absolutely continuous with
respect to Lebesgue measure.
\end{proposition}

\textbf{Proof} See (\cite{C}, exercise 11, pg. 159) or \cite{Si}; for a
generalization of this result using different methods, see \cite{Es}.

We now go back to Theorem \ref{sharp}. Let 
\begin{equation}
\lambda ^{\pm }=\pm 2\sqrt{1-\frac{v^{2}}{v_{c}^{2}}}  \label{lambda}
\end{equation}%
under the condition%
\begin{equation}
0<v<v_{c}=2\sqrt{\beta -1}  \label{lambda1}
\end{equation}%
so that%
\begin{equation}
0<\lambda ^{+}<2~.  \label{lambda2}
\end{equation}

We are now ready to state our main result:

\begin{theorem}
\label{main}Let $J_{\theta }^{(2)}$ be defined by (\ref{J2}) and let%
\begin{equation}
v^{2}<a\left( \sqrt{\beta }-1\right) <v_{c}^{2}~  \label{vv}
\end{equation}%
with $a<4$. Then, for almost every $\left( \omega ^{1},\omega ^{2},\theta
\right) $ with respect to $\nu \times \nu \times l$,

\begin{enumerate}
\item[a.] there exist $\tilde{\lambda}^{\pm }$ with $\tilde{\lambda}^{+}=-%
\tilde{\lambda}^{-}$ and 
\begin{subequations}
\label{til}
\begin{equation}
0<\tilde{\lambda}^{+}<\lambda ^{+}  \label{tilde}
\end{equation}%
such that%
\begin{equation}
\left( \tilde{\lambda}^{-}(1+\theta ),\tilde{\lambda}^{+}(1+\theta )\right)
\subset \sigma _{\mathrm{ac}}\left( J_{\theta }^{(2)}\right)  \label{tilde1}
\end{equation}

\item[b.] 
\begin{equation}
\left[- 2(1+\theta ),\lambda ^{-}(1+\theta )\right) \cup \left( \lambda
^{+}(1+\theta ),2(1+\theta )\right] \subset \sigma _{\mathrm{pp}}\left(
J_{\theta }^{(2)}\right)  \label{tilde2}
\end{equation}

\item[c.] 
\begin{equation}
\sigma _{\mathrm{sc}}\left( J_{\theta }^{(2)}\right) \cap \left( \lambda
^{-}(1+\theta ),\lambda ^{+}(1+\theta )\right)  \label{tilde3}
\end{equation}
may, or may not, be an empty set.
\end{subequations}
\end{enumerate}
\end{theorem}

\noindent \textit{Proof.} We first choose $I_{0}$ in Corollary \ref{corl}
such that 
\begin{equation}
I_{0}=\left[ -\tilde{\lambda}^{+},\tilde{\lambda}^{+}\right]  \label{I0}
\end{equation}%
and%
\begin{equation}  \label{alphac}
\alpha =\min_{i:\tilde{A}_{\varepsilon }^{i}\cap I_{0}\neq \emptyset }\alpha
(\lambda _{i})-\varepsilon >\frac{1}{2}~.
\end{equation}
The inequalities (\ref{tilde}) and (\ref{alphac}) are established in
Appendix \ref{PAR} (Proposition \ref{param}) for any choice of parameters $p$%
, $\beta $ satisfying (\ref{vv}) and $\varepsilon $ depending on $p$, $\beta 
$ and $a$.

Coming back to (\ref{fsc}), by polarization we need only consider $\varphi
_{1}=\psi _{1}\in D_{1}$, $\varphi _{2}=\psi _{2}\in D_{2}$ and,
accordingly, with (\ref{I0}) and (\ref{alphac}), we define%
\begin{equation}
f_{\mathrm{sc}}^{i}(s):=\int_{I_{0}}e^{-is\lambda }d\mu _{\varphi _{i}}^{%
\mathrm{sc}}(\lambda )~,\qquad i=1,2  \label{fsci}
\end{equation}%
in (\ref{three}).

Let%
\begin{equation}
I_{i}(T):=\int_{0}^{T}\left\vert f_{\mathrm{sc}}^{i}(s)\right\vert ^{2}ds
\label{Ii}
\end{equation}%
By Strichartz' theorem \cite{Str} (see also Theorem 2.5 of \cite{L1}, for a
slick proof) and (\ref{I0})%
\begin{equation}
I_{i}(T)\leq C_{i}T^{1-\alpha }\leq CT^{1-\alpha }  \label{IiT}
\end{equation}%
for $0<C_{i}<\infty $, $i=1,2$, $T$--independent constants and $C=\max
(C_{1},C_{2})$. By (\ref{fsci}) and (\ref{IiT}) and a change of variable, we
have%
\begin{equation*}
\int_{0}^{1}\left\vert f_{\mathrm{sc}}^{2}(\theta t)\right\vert ^{2}d\theta =%
\frac{1}{t}I_{2}(t)\leq Ct^{-\alpha }~
\end{equation*}%
which implies%
\begin{equation}
\int_{1}^{T}dt\left\vert f_{\mathrm{sc}}^{1}(t)\right\vert
^{2}\int_{0}^{1}\left\vert f_{\mathrm{sc}}^{2}(\theta t)\right\vert
^{2}d\theta \leq C\int_{1}^{T}\left\vert f_{\mathrm{sc}}^{1}(t)\right\vert
^{2}t^{-\alpha }dt~.  \label{intff}
\end{equation}%
We now perform an integration by parts on the r.h.s. of (\ref{intff})%
\begin{equation}
\int_{1}^{T}\left\vert f_{\mathrm{sc}}^{1}(t)\right\vert ^{2}t^{-\alpha
}dt=\left. I_{1}(t)t^{-\alpha }\right\vert _{1}^{T}+\alpha
\int_{1}^{T}dtI_{1}(t)t^{-\alpha -1}  \label{intf}
\end{equation}%
By (\ref{intff}), (\ref{intf}) and Fubini's theorem ($T\geq 1$)%
\begin{eqnarray}
\int_{0}^{1}d\theta \int_{1}^{T}\left\vert f_{\mathrm{sc}}^{1}(t)\right\vert
^{2}\left\vert f_{\mathrm{sc}}^{2}(\theta t)\right\vert ^{2}dt &\leq
&CT^{1-2\alpha }+\alpha C\int_{1}^{T}dtt^{-2\alpha }  \notag \\
&\leq &C\frac{1}{2\alpha -1}\left( \alpha -(1-\alpha )T^{1-2\alpha }\right)
~.  \label{intff1}
\end{eqnarray}%
By (\ref{alphac}) and (\ref{intff1}), the limit%
\begin{equation*}
\int_{0}^{1}d\theta \int_{0}^{\infty }\left\vert f_{\mathrm{sc}%
}^{1}(t)\right\vert ^{2}\left\vert f_{\mathrm{sc}}^{2}(\theta t)\right\vert
^{2}dt=\lim_{T\rightarrow \infty }\int_{0}^{1}d\theta \int_{0}^{T}\left\vert
f_{\mathrm{sc}}^{1}(t)\right\vert ^{2}\left\vert f_{\mathrm{sc}}^{2}(\theta
t)\right\vert ^{2}dt~
\end{equation*}%
exists, is finite and%
\begin{equation}
\int_{0}^{\infty }\left\vert f_{\mathrm{sc}}^{1}(t)\right\vert
^{2}\left\vert f_{\mathrm{sc}}^{2}(\theta t)\right\vert ^{2}dt<\infty
\label{intff2}
\end{equation}%
for a.e. $\theta \in \left[ 0,1\right] $. By Ichinose's theorem \cite{I}
(actually, Theorem VIII.33 of \cite{RS}, for $A_{k}$ bounded, and its
Corollary, pgs. 300 and 301, suffice) and (\ref{J2}), the spectrum of $%
J_{\theta }^{(2)}$ is the arithmetic sum of the spectrum of $J_{\phi
}^{\omega ^{1}}$ and $\theta J_{\phi }^{\omega ^{2}}$. Together with Theorem %
\ref{sharp}, Proposition \ref{ac} and (\ref{intff2}) this proves (\ref%
{tilde1}).

In order to prove (\ref{tilde2}), we need only consider $\rho _{1}=\chi
_{1}\in \mathcal{H}_{\mathrm{pp}}$ and $\rho _{2}=\chi _{2}\in \mathcal{H}_{%
\mathrm{pp}}$ with $f_{\mathrm{pp}}^{i}$ in (\ref{fpp}) defined accordingly.
By Theorem 5.6 of \cite{Kat}, $\mathbb{R}\ni t\longmapsto f_{\mathrm{pp}%
}^{i}(t)$ is an almost periodic function on $\mathbb{R}$, i.e., $f_{\mathrm{%
pp}}^{i}\in AP\left( \mathbb{R}\right) $ (see \cite{Kat}, Definitions 5.1
and 5.2) and, therefore, (see \ref{three3})%
\begin{equation*}
k(t,\theta )=f_{\mathrm{pp}}^{1}(t)f_{\mathrm{pp}}^{2}(\theta t)
\end{equation*}%
belongs to $AP\left( \mathbb{R}\right) $ by Theorem 5. of \cite{Kat} and,
again by Theorem 5.6 of \cite{Kat}, $\mu $ defined by%
\begin{equation*}
\mu =\mu _{\rho _{1}}^{\mathrm{pp}}\ast \tilde{\mu}_{\rho _{2}}^{\mathrm{pp}}
\end{equation*}%
where $\tilde{\mu}_{\rho _{2}}^{\mathrm{pp}}(\lambda )=\mu _{\rho _{2}}^{%
\mathrm{pp}}(\lambda /\theta )$, $\theta \neq 0$, is pure point. Together
with Ichinose's theorem and Theorem \ref{sharp}, this proves (\ref{tilde2}).

By the definition analogous to (\ref{conv}) it follows that 
\begin{equation*}
\mu _{\rho _{1}}^{\mathrm{pp}}\ast \tilde{\mu}_{\varphi _{2}}^{\mathrm{sc}%
}\left( \{\lambda \}\right) =\mu _{\varphi _{1}}^{\mathrm{sc}}\ast \tilde{\mu%
}_{\rho _{2}}^{\mathrm{pp}}\left( \{\lambda \}\right) =0
\end{equation*}%
for any singleton $\left\{ \lambda \right\} $. Hence, by Ichinose's theorem
and Theorem \ref{sharp}, the spectrum of $J_{\theta }^{(2)}$ restricted to $%
\left[ (1+\theta )\lambda ^{-},(1+\theta )\lambda ^{+}\right] $ is
necessarily continuous -- but may be singular continuous -- showing part 
\textbf{c.} and concluding the proof of Theorem \ref{main}.

\hfill $\Box $


\begin{remark}
Some of the ideas used in the proof of Theorem \ref{main} have also employed
by Kahane and Salem \cite{KS1,KS2} in more specific contexts. We refer in
particular to \cite{KS2} for the general crucial method of interpolating the
sets $\left\{ \xi _{k}\right\} $ of dissection ratios of Cantor sets by
convex combinations%
\begin{equation*}
\xi _{k}=a_{k}(1-\zeta _{k})+\xi \zeta _{k}
\end{equation*}%
with $\zeta \equiv \left( \zeta _{1},\ldots ,\zeta _{k},\ldots \right) $ in
the unit hypercube, and then proving that F.S. transform of the
corresponding s.c. measure tends to zero at infinity for a.e. $\zeta $ (Th%
\'{e}or\`{e}me III of \cite{KS2}, pg. 106). In our case the parameter $%
\theta $ (the analog of $\zeta $) appears in (\ref{J2}), and the F.S.
transform of the corresponding measure is $L^{2}$ for a.e. $\theta \in \left[
0,1\right] ,$ which implies that it tends to zero at infinity by the
Riemann--Lebesgue lemma.
\end{remark}

\begin{remark}
The a.c. part of the spectrum of $J_{\theta }^{(2)}$ is not, of course,
promoted by the randomness on the \textquotedblleft bump" positions. It
makes, however, the Hausdorff dimension of the spectral measures $\mu
_{\varphi _{1}}^{\mathrm{sc}}$ and $\tilde{\mu}_{\varphi _{2}}^{\mathrm{sc}}$
and, consequently, the intervals $I_{0}$ and $I$ appearing in Theorems \ref%
{main} and \ref{sharp}, be determined exactly. Items $a.$and $b.$ of Theorem %
\ref{main} thus hold for a bidimensional model (\ref{J2}) with the $J_{\phi
}^{\omega _{i}}$ replaced by deterministic sparse models studied in \cite%
{MWGA} since their local Hausdorff dimension may be determined as accurately
as one wishes, provided the sparse parameter $\beta $ is large enough. The
p.p. part of the spectrum cannot, however, be established except for the
random model (see comment after Theorem 2.3 of \cite{CMW1} and Remark 5.9.1
of \cite{MWGA}).
\end{remark}

\begin{remark}
It is important to employ our version of Zlato\v{s}'s theorem (Theorem 2.4
of \cite{CMW1}), which shows the purity of the p.p. spectrum. For, in case
that the p.p. spectrum contains admixture of s.c. spectrum, the latter may,
by convolution, generate an a.c. part in $J_{\theta }^{(2)}$. Since a
(possibly dense) p.p. superposition to the a.c. spectrum of $J_{\theta
}^{(2)}$ cannot be excluded in Theorem 2.1 (originated e.g. from the
convolution of two -- again possibly dense -- p.p. spectra which may be
superposed to the s.c. spectrum of Theorem 2.4 of \cite{CMW1}), we would, in
this special case, have no transition at all in the spectral type from one
region to another.
\end{remark}

\begin{remark}
\label{resonance}In the special case of exactly self--similar spectral
measures $\mu $ and $\mu _{\theta }$ ($\mu _{\theta }(\lambda )=\mu (\lambda
/\theta )$), a theorem of X. Hu and S. J. Taylor \cite{HT} implies that
their convolution is a.e. $\theta \in \left[ 0,1\right] $ absolutely
continuous. This fact has been used by Bellissard and Schulz--Baldes \cite%
{BS} to construct the first models in $d\geq 2$ dimensions with $a.c.$
spectrum and subdiffusive quantum transport (thought to describe properties
of quasicrystals) -- see their theorem in \cite{BS} and a previous remark
that it cannot be true for all $\theta $ due to resonance phenomena; see
also \cite{PS}. It is to be remarked that exact self--similarity is a rare
property. In particular, Combes and Mantica \cite{CM} proved that this
property does not hold for sparse models, such as ours (see Theorem 2 of 
\cite{CM}).
\end{remark}

\begin{remark}
It is clear that the proof of Theorem \ref{main} generalizes to dimensions $%
d>2$, for even a wider range of parameter values, since the corresponding
condition on the r.h.s. of (\ref{alphac}), given by $\alpha >1/d$, becomes
successively weaker for increasing $d$.
\end{remark}

\begin{remark}
\label{decay}We have not proved pointwise decay of the F.S. transform $\hat{%
\mu}$ of the spectral measure $\mu $ of $J_{\phi }^{\omega };$ i.e., a bound
of the form%
\begin{equation}
\left\vert \widehat{|f|^{2}\mu }(t)\right\vert \leq C_{f}t^{-\alpha /2}~
\label{pointwise}
\end{equation}%
for $C_{0}^{\infty }\left( [-2,2]\right) $ functions $f.$ Indeed, such a
bound (\ref{pointwise}) has never been proved except for classes of sparse
models with superexponential sparsity, for which the spectrum is purely s.c.
and the Hausdorff dimension equal to one; in this case, (\ref{pointwise})
assumed the form: $\forall \varepsilon >0$, $\exists $ $0<C_{\varepsilon
}<\infty $ such that 
\begin{equation}
\left\vert \widehat{|f|^{2}\mu }(t)\right\vert \leq C_{f,\varepsilon
}t^{-1/2+\varepsilon }  \label{ptw}
\end{equation}%
(see \cite{S2,KR,CMW3}). It is a challenging open problem to prove (\ref{ptw}%
) for the present model, with $1/2$ replaced by $\alpha /2$ on the r.h.s.
with $\alpha $ being the local Hausdorff dimension.
\end{remark}

\section{Conclusions and Open Problems\label{COP}}

\setcounter{equation}{0} \setcounter{theorem}{0}

Our main result (Theorem \ref{main}) realizes part of the program set by
Molchanov in dimensions $d\geq 2$. See also the discussion in Chap.5 of \cite%
{DK}.

Concerning possible physical applications, it seems natural to expect that
the present model might pave the way for a good qualitative description of
the Anderson transition in lightly doped semiconductors, which, in fact,
takes place for $d\geq 2$! (see Chap. 2.2 of \cite{SE}). We say
\textquotedblleft pave the way" because the present form of the model is not
adequate for a physical description for at least two reasons -- but we argue
that both objections may be eliminated by considering a truly $d$%
--dimensional model.

The first reason is, of course, that exponential sparsity (\ref{sparse}) is
too severe, and not physically reasonable. It must be recalled, however,
that the separable model does not take account of dimensionality in a proper
way. For instance, for the usual one--dimensional model (see e.g. \cite%
{GMP,KS}), supposedly adequate to describe heavily doped semiconductors, the
three dimensional version (analogous to (\ref{J2})) also yields purely p.p.
spectrum, by the same proof of Theorem \ref{main}, in complete disagreement
with the expected transition (see also Section \ref{IS}). However,
\textquotedblleft truly" three dimensional sparse models may drastically
change, in (\ref{a}), the cardinality of $\left\{ i:\left\vert
a_{i}\right\vert \leq R\right\} $ from $O\left( \log R\right) $ to $%
O(R^{d-\varepsilon })$ in dimension $d$, for some $\varepsilon >0$, which is
still compatible with (\ref{a}), changing, at the same time, the conditions
on the sparsity for the existence of the transition.

The second reason is that, in one dimension, exponential sparsity (\ref%
{sparse}) is critical for the existence of transition: there is no
transition (at least for $0<p_{a_{k}}<1$) either for subexponential or for
superexponential sparsity (See Section \ref{IS}, for discussion and
references). Again, for \textquotedblleft truly" $d\geq 2$ dimensional
systems we expect this to change, implying a wider region in the sparsity
parameter for which a transition takes place.

As in the Bethe lattice case treated by \cite{Kl}, the sharpness of the
transition, i.e., the existence of a mobility edge, was not proved for the
present model. The recent surprising work of Aizenman and Warzel \cite{AW1}
proves that no mobility edge occurs in the Bethe lattice at weak disorder.
Similarly to the Bethe lattice, our separable model has no loops, but it is
certainly a constituent part of the full model in $d$ dimensions (for light
doping, as conjectured above). The general character of the arguments used
in Theorem \ref{main} to establish the existence of a.c. spectrum, which we
commented upon at the end of the introduction, suggests that the
intermediate region might be more accessible to analysis than the the Bethe
lattice, but this remains as a challenging open problem. On the other hand,
it is rewarding that already the separable model displays a dramatic
\textquotedblleft kinematic" effect of the dimensionality: for $d\geq 2$ the
transition becomes truly Anderson--like, i.e., from a.c. to p.p. spectrum.
The a.c. spectrum is the one which most closely corresponds to the
physicist's picture of \textquotedblleft delocalized states"; indeed, the
s.c. spectrum has quite different properties, both dynamic \cite{Si} and for
the point of view of perturbations (see e.g. \cite{SiWo,H2}).

Finally, it is clear that, besides the intermediate region mentioned above,
Theorem \ref{main} leaves much room for improvement. Elimination of the set
of zero Lebesgue measure in the s.c.part of the spectrum would be a
significant improvement, as a well as clarification of which alternative
holds in item \textbf{c.} of Theorem \ref{main}.

\appendix

\section{The Choice of Parameters\label{PAR}}

\setcounter{equation}{0} \setcounter{theorem}{0}

\begin{proposition}
\label{param}Let $p\in \left( 0,1\right) $ and $\beta \geq 2$ be chosen so
that (\ref{vv}) holds for some $a<4$. Then, there exists $\varepsilon
_{0}=\varepsilon _{0}(p,\beta ,a)>0$ such that (\ref{tilde}) and (\ref%
{alphac}), with $I_{0}$ given by (\ref{I0}), are satisfied for any $%
0<\varepsilon <\varepsilon _{0}$.
\end{proposition}

\noindent \textit{Proof.} With the definitions (\ref{r}) of $r(\lambda )$
and (\ref{delta-bar}), let $I_{0}=\left[ \tilde{\lambda}^{-},\tilde{\lambda}%
^{+}\right] $, $\tilde{\lambda}^{-}=-\tilde{\lambda}^{+}$, be defined by%
\begin{equation}
r(\tilde{\lambda}^{+}+\bar{\delta}_{\varepsilon })=r^{\ast }~  \label{b1}
\end{equation}%
for certain $r^{\ast }$ satisfying 
\begin{equation*}
1+\frac{v^{2}}{4-\bar{\delta}_{\varepsilon }^{2}}<r^{\ast }<\sqrt{\beta }~.
\end{equation*}%
By the first inequality there exists $\tilde{\lambda}^{+}>0$ which solves (%
\ref{b1}). Note that $r(\lambda )$ is monotone increasing for $\lambda \in
\left( 0,2\right) $. Under the condition (\ref{vv}), with $a<4$ fixed,%
\begin{equation*}
1+\frac{v^{2}}{4-\bar{\delta}_{\varepsilon }^{2}}<1+\frac{a}{4-\bar{\delta}%
_{\varepsilon }^{2}}\left( \sqrt{\beta }-1\right) <\sqrt{\beta }
\end{equation*}%
by (\ref{delta-bar}), provided $\varepsilon <\varepsilon _{1}$ for some $%
\varepsilon _{1}=\varepsilon _{1}(p,\beta ,a)>0$. So, $r^{\ast }$ is well
defined and 
\begin{equation*}
0<\tilde{\lambda}^{+}<\lambda ^{+}
\end{equation*}%
by (\ref{lambda1}), monotonicity of $r(\lambda )$ and $r(\tilde{\lambda}%
^{+})<\sqrt{\beta }<\beta =r\left( \lambda ^{+}\right) $, for $\beta \geq 2$.

In addition, it follows by (\ref{b1}) and equations (\ref{d}a-f) that $%
\left\vert \lambda _{i}\right\vert \leq \tilde{\lambda}^{+}+\bar{\delta}%
_{\varepsilon }$ holds for every $i$ such that $\tilde{A}_{\varepsilon
}^{i}\cap I_{0}\neq \emptyset $ and, by definition (\ref{alpha-def}), (\ref%
{alpha}) and the monotone behavior of $r(\lambda )$, 
\begin{equation*}
\alpha =\min_{i:\tilde{A}_{\varepsilon }^{i}\cap I_{0}\neq \emptyset }\alpha
(\lambda _{i})-\varepsilon \geq 1-\frac{\ln r^{\ast }}{\ln \beta }%
-\varepsilon >\frac{1}{2}
\end{equation*}%
provided $\varepsilon <\varepsilon _{0}$ with $\varepsilon _{0}=\min
(\varepsilon _{1},\ln \left( \sqrt{\beta }/r^{\ast }\right) /\ln \beta )>0$,
establishing (\ref{alphac}). This concludes the proof of the proposition.

\hfill $\Box $

\begin{center}
\textbf{Acknowledgements}
\end{center}

DHUM thanks Gordon Slade and David Brydges for their hospitality at UBC. We
thank the referee for the recommendation of research directions and
references.

\end{document}